\documentclass[preprint,12pt]{elsarticle}
\usepackage[utf8]{inputenc} 
\usepackage[T1]{fontenc}    
\usepackage{url}            
\usepackage{booktabs}       
\usepackage{amsfonts}       
\usepackage{nicefrac}       
\usepackage{microtype}      
\usepackage{lipsum}
\usepackage{times}
\usepackage{amsmath}
\usepackage{graphicx}
\usepackage{multicol}
\usepackage{longtable}
\usepackage[refpages]{gloss}
\usepackage{float}
\usepackage{anysize}
\usepackage{appendix}
\usepackage{lscape} 
\usepackage{pdflscape}
\usepackage{multirow}
\usepackage{listings}
\usepackage{color}
\usepackage{setspace}
\usepackage{enumerate} 
\usepackage[export]{adjustbox}
\usepackage{epsfig}
\usepackage{graphicx}
\usepackage{caption}
\usepackage{subcaption}
\usepackage[export]{adjustbox}
\usepackage{wrapfig}
\usepackage{url}
\usepackage{float}

\usepackage{natbib}
\usepackage{algpseudocode}
\usepackage{algorithm}
\usepackage{rotating}
\usepackage{multirow}
\usepackage{array}
\usepackage{graphicx}
\usepackage{amssymb}
\usepackage{fixltx2e}
\usepackage{emojione}
\usepackage{CJKutf8}
\usepackage{scrextend}
\usepackage[colorlinks]{hyperref}

\makeatletter
\def\ps@pprintTitle{%
  \let\@oddhead\@empty
  \let\@evenhead\@empty
  \let\@oddfoot\@empty
  \let\@evenfoot\@oddfoot
}
\makeatother

\begin{document}

\begin{frontmatter}

%% Title, authors and addresses

\title{FATE in AI: Towards Algorithmic Inclusivity and Accessibility}

\author{Isa Inuwa-Dutse}
\address{University of Huddersfield, UK\\
\texttt{i.inuwa-dutse@hud.ac.uk}}
%\address[label2]{University of Hertfordshire}

\begin{abstract} 

Artificial Intelligence (AI) is at the forefront of modern technology, and its effects are felt in many areas of society. To prevent algorithmic disparities, fairness, accountability, transparency, and ethics (FATE) in AI are being implemented. However, the current discourse on these issues is largely dominated by more economically developed countries (MEDC), leaving out local knowledge, cultural pluralism, and global fairness. This study aims to address this gap by examining FATE-related desiderata, particularly transparency and ethics, in areas of the global South that are underserved by AI. A user study ($n=43$) and a participatory session ($n=30$) were conducted to achieve this goal. The results showed that AI models can encode bias and amplify stereotypes. To promote inclusivity, a community-led strategy is proposed to collect and curate representative data for responsible AI design. This will enable the affected community or individuals to monitor the increasing use of AI-powered systems. Additionally, recommendations based on public input are provided to ensure that AI adheres to social values and context-specific FATE needs.

\end{abstract}


\end{frontmatter}

%%
%% Start line numbering here if you want
%%
%\linenumbers

%% main text
% INTRODUCTION:\input{p1-introduction.tex}
% BACKGROUND: ... and related work\input{p2-background.tex}
% MICROCOSM DETECTION \input{p3-microcosm-detection.tex}
% EXPERIMENTATION/DISCUSSION\input{p4-experimentation.tex}
% CONCLUSION\input{p5-conclusion.tex}

\section{Introduction}
\label{sec:introduction} 

The increasing use of Artificial Intelligence (AI) technology in various domains has the potential to cause individual and social harm. Examples of bias and discrimination in AI applications include court decisions \cite{compass2016}, job hiring \cite{jobhiringAI}, online ads \cite{adshowingAI}, and many other areas prone to bias \cite{schmidt2019introduction}. These algorithmic decisions have economic and personal implications for individuals. Therefore, Fairness, Accountability, Transparency and Ethics (FATE) in AI must be properly regulated for responsible use cases \cite{shin2020beyond,ferrario2020ai}, particularly in high-stakes domains \cite{compass2016,adadi2018peeking,aibias2020news,makingprofiling,carton2016identifying,eubanks2018automating,grgic2019human}. Studies have shown that machine learning models can discriminate based on race and gender \cite{bolukbasi2016man,buolamwini2018gender,dastin2018amazon}. FATE in AI is intended to address the social issues caused by digital systems, but the current discourse is largely shaped by more economically developed countries (MEDC), raising concerns about neglecting local knowledge, cultural pluralism, and global fairness \cite{jobin2019global}. As AI systems become more integrated into various products \cite{makingprofiling,carton2016identifying,gebru2019oxford,grgic2019human,sturm2016interpretable,Rago:2020}, they are a major driver of the fourth industrial revolution (4IR) and transformation \cite{4ir2016}. Therefore, it is essential to understand the FATE-related needs of different communities, as AI affects a wide range of people. Ensuring effective transparency cannot be a one-size-fits-all approach \cite{doshi2017towards}, as this could disproportionately affect different communities \cite{jobin2019global,sambasivan2021re}. To this end, more contextualised and interdisciplinary research is needed to inform algorithmic fairness and transparency \cite{friedler2019comparative,boykin2021opportunities,ribera2019can}. Additionally, diversity and sociodemographics must be taken into account when designing and governing algorithms that affect the public \cite{fazelpour2022diversity}. The most effective approach is to involve the affected public and AI developers to incorporate community-specific FATE needs. AI practitioners must adhere to social values to ensure responsible AI for the public good \cite{selbst2019fairness}, and focus on more representative values of the community affected by AI \cite{jakesch2022different}. Through cooperative, inclusive, and community-led design of AI applications, algorithmic disparities can be effectively addressed, and relevant stakeholders within the community can ensure better policing of AI operations.

%update with research aim/questions: 
%\footnote{the following contributions will be presented as part of non-archival submission during EAAMO 23 conference}
This study aims to explore the areas of AI that are under-served in terms of responsibility and accountability. It seeks to evaluate the effectiveness of transparency methods in relation to FATE-related issues, as well as to explore ways in which local communities can be involved in the design and development of AI systems that affect them. Additionally, it will provide useful insights and recommendations to stimulate action towards representative and responsible AI. To this end, a community of 73 online users in Nigeria from the Global South was chosen as a case study to examine the public's views on FATE in AI. Nigeria was selected due to its large population and the increasing use of AI-powered products and services. In addition, the country is ranked 8th in the world for Internet users \cite{factbook2022}, indicating a growing AI workforce in Africa. 

The paper is divided into the following sections. Section~\ref{sec:background} provides the background and related studies. Section~\ref{sec:study-design} outlines the method and details of user studies. Sections~\ref{sec:participants-response} and ~\ref{sec:discussion} present the results and a discussion, respectively. Finally, Section~\ref{sec:conclusion} concludes the study.

%To bridge the accessibility and inclusivity gap, the study\footnote{part of the result in this study was presented as a poster \cite{inuwaai}} and non-archival in EAAMO13.... TO BE INCLUDED AFTER A SUCCESSFUL DECISION 

\section{Background} 
\label{sec:background} 
This research draws from socio-technical fields to investigate FATE in AI matters. We will start by looking at the application areas of AI, followed by an overview of pertinent literature from the fields of responsible AI and human-computer interaction (HCI). The uniqueness of our strategy is that it involves the communities affected by the development of responsible and inclusive AI systems.

\subsection{AI Application Areas}  
\label{sec:application-areas}
The utilisation of Artificial Intelligence (AI) in Africa is becoming increasingly widespread, with applications in a variety of areas such as governance, healthcare, agriculture, financial institutions, transportation, education, journalism, public and private engagements. In healthcare, AI is used for automated diagnostics (for example, minoHealth\footnote{\url{https://www.minohealth.ai/}}), pharmacy management systems (for example, RxAll\footnote{\url{https://rxall.net/about/}}), and medical imaging \cite{botwe2021integration,antwi2021artificial}. AI is also used in journalism for newsgathering, production, and distribution processes \cite{kothari2022artificial}, and in public administrations to explore the role and implications of AI \cite{plantinga2022digital}. 
Financial institutions are utilising AI tools such as Debtors\footnote{\url{https://www.debtorsafrica.com/Home/AboutUs}} and Nomba\footnote{\url{https://nomba.com/about-us}} to reduce discrimination against black people in identity verification. In addition, research has been conducted on the malicious use of AI in Sub-Saharan African countries \cite{pantserev2022malicious} and how AI could promote sustainability and tackle climate challenges in Africa \cite{rutenberg2021use}. Despite the potential of AI to address various social issues, the development and deployment of AI systems requires a significant amount of energy \cite{ad2020first, nishant2020artificial}. 

%\subsection{Societal Impact of AI} %\label{sec:societal-impact} 
The use of AI in low- and middle-income countries (LMICs) will have positive and negative effects \cite{smith2018artificial}. These tools and platforms often involve the collection and storage of sensitive data about individuals, which raises ethical issues. Previous studies have shown that policy responses to mitigate and prevent bias results are not keeping up with the rapid deployment of AI technologies in Africa \cite{hao2019future,gwagwa2020artificial,neupane2021artificial,ade2023artificial,eke2023towards}. With the rapid development of AI, it is reasonable to ask what would happen if Artificial General Intelligence (AGI) is achieved, giving AI powerful reasoning and task-accomplishment capabilities. In the AGI era, will AI be able to determine what is right and wrong if its intelligence surpasses that of the human user? What would be the impact or risk of jobs being replaced by AI due to automation in areas with already high unemployment and other social problems? Addressing current AI challenges in underserved communities will help address the most pressing AI issues. 

\subsection{Algorithmic Experience and Fairness}
\label{sec:algorithmic-fairness} 
Algorithmic decisions can lead to the reproduction or intensification of disparities for a variety of reasons. For example, discrimination can be inherent due to the data used to train the AI model, which may have been biased and discriminatory in the first place \cite{dodge2019explaining}. As mentioned above, algorithms can also promote a form of discrimination and stereotyping. Examples of this include the COMPAS system to predict recidivism \cite{compass2016}, Amazon's hiring process that favoured male applicants over females \cite{jobhiringAI}, and Google's job ad algorithm that showed jobs paid higher for men than women \cite{adshowingAI}. Consequently, fairness, accountability, transparency, and ethics in AI are aimed at developing and ensuring responsible AI that incorporates moral behaviour and avoids encoding bias in AI decisions \cite{kulynych2020pots}. Ethics is about making choices based on concepts of right and wrong, duty and obligation. Therefore, it is possible to create a hierarchy of goals that embody ethical principles in digital systems \cite{waldrop1987question}.

\paragraph{Regulatory Frameworks} 
Legislation, relevant policies, and positive action are essential for addressing ethical issues and algorithmic disparities \cite{bynum2021disaggregated}. The Equality Act 2010 in the UK \footnote{\url{shorturl.at/cqr19}} uses positive action to reduce the imbalance of opportunity for individuals from under-represented communities. Algorithmic fairness is also seen through the lens of positive action to promote equal representation \cite{thomas2021algorithmic}. Additionally, ethical frameworks such as UNESCO's recommendation on the Ethics of AI \cite{unesco2021}, the World Economic Forum's blueprint for inclusive AI \cite{wef2022}, and the Organisation for Economic Co-operation and Development (OECD) \cite{yeung2020recommendation} have been proposed to address the human rights implications of AI systems. Regulatory bodies such as the steering group of the European AI Alliance \cite{europeancomm2022} and the African Working Group on AI \cite{/content/publication/bb167041-en} are also in place to monitor AI operations and protect human rights \cite{OHCHR2021}. To ensure fairness, it is important to consider how inclusive the AI design and development pipeline is in terms of demographics and local context \cite{vincent2021data}. Data leverage \cite{arrieta2018should} can be used to empower the public to influence the implementation process and tackle algorithmic unfairness \cite{eubanks2018automating,gebru2019oxford,kulynych2020pots,abebe2020roles}.
%\subsection{Algorithmic Experience} %\label{algorithmic-experience}
\paragraph{Algorithmic Transparency and Fairness}
The importance of transparency in Artificial Intelligence (AI) systems is widely recognised as a necessary condition for upholding fundamental human rights and ethical principles \cite{ad2020first}. However, there is a wide range of interpretations, justifications, applications, and methods for achieving transparency \cite{jobin2019global}. Explainable AI (XAI) provides explanations to humans to help them better understand AI systems \cite{adadi2018peeking,bhatt2020explainable}, and can be used to identify and address fairness issues \cite{dodge2019explaining,bansal2021does,buccinca2021trust}. Explanations should be tailored to the individual requesting them, and not necessarily be designed to make the decision process understandable to the end-user \cite{andras2018trusting}. This shift from developer-driven explanations \cite{ferreira2020people} to those that satisfy users' curiosity and increase their knowledge about the technology \cite{hoffman2018metrics} requires a user-centric approach. Previous research has highlighted the need for contextualised and interdisciplinary studies to identify best practices for AI \cite{friedler2019comparative,boykin2021opportunities,ribera2019can}. This study emphasises the importance of providing appropriate transparency to the public, so that they can be informed about the basis of AI decisions, as they have the right to know.

\subsection{Data and Community Voice} 
As the population of many countries in the Global South grows, they are generating a large amount of data that is providing value and competitive advantage to numerous technology companies. AI technologies are being used in new user groups, applications, datasets, and regulations \cite{sambasivan2018toward}. However, ethical, privacy, and data protection issues have been raised due to the fact that AI systems require a great deal of data for training and functioning properly \cite{smith2018artificial,adams2020introducing}. The dominance of Western viewpoints has been attributed to the prevalence of data, measurement scales, legal and philosophical dimensions \cite{sambasivan2021re}, which has led to AI being shaped by its originating contexts in Western nations \cite{sambasivan2018toward}. This has been seen as a new form of dominance, with the commercialisation and weaponisation of data facilitated by local and foreign AI models being referred to as data colonialism \cite{coleman2018digital,pilling2019tech,couldry2019data,couldry2019data}. Relying on foreign data generation and processing tools has been criticised due to privacy and data protection concerns \cite{foreignai2020}. Techno-colonialism has been argued to be a new way of transferring technology and its values and norms from more economically developed countries to developing ones \cite{ugar2023fourth}. To ensure responsible AI that is in line with the affected community's values, access to representative and quality data is essential. Therefore, creating transparent, trustworthy, and human-centric AI that is in agreement with context-specific norms requires input from diverse stakeholders, and high regard for diversity and sociodemographics should be taken into account in the design and governance of algorithms that affect the public \cite{fazelpour2022diversity}. Not incorporating local context or sociodemographic factors could lead to (in)advertent or (un)intended discrimination and amplifying biases present in data \cite{bolukbasi2016man}. Interdisciplinary approaches that are beneficial to various stakeholders' desiderata and for evaluation purposes should be incorporated \cite{langer2021we,mohseni2021multidisciplinary,ribera2019can}. Within the HCI research community, approaches have been put forward to find better ways of improving the transparency of digital systems \cite{cheng2019explaining,ribera2019can,shin2021effects,kroeger2022social}. Explanation interfaces that inform users about the inner workings of the algorithm and interactivity are effective for algorithmic intuitiveness and improved transparency \cite{cheng2019explaining}. To better improve inclusivity and leverage AI's capability, the involvement of various stakeholders to sketch out plans for AI applications that recognise local context or needs is essential \cite{hsu2022empowering}. This study is focused on how the communities can lend their voice and influence FATE in AI discourse, in line with existing initiatives that support AI developers to factor in community-specific needs \cite{orife2020masakhane,gebru2021datasheets}. 

\subsection{AI and Policy Response} 
\label{sec:ai-and-policy-response} 
As Artificial Intelligence (AI) becomes increasingly pervasive in our lives, there is growing concern about the potential negative impacts it may have on ethics and human rights. This has led to calls for laws, policies and guidelines to regulate its development and application \cite{brandusescu2017artificial,smith2018artificial,owoyemi2020artificial,gwagwa2020artificial,OHCHR2021,policybrief2021,wakunuma2022responsible}. Previous studies have noted the rapid deployment of AI technologies in Africa, but policy responses to mitigate and prevent biased outcomes are lagging \cite{hao2019future,onuoha2019ai,gwagwa2020artificial,neupane2021artificial,ade2023artificial,stahl2023ai,hassan2023governing}. To this end, initiatives have been put in place to contain the technology within the agreed parameters. For example, the OECD adopted the Recommendation on AI, which is intended to promote innovation and trust in the technology \cite{yeung2020recommendation}. In addition, many African countries have established measures to govern AI. For instance, the African Working Group on AI has been established \cite{/content/publication/bb167041-en}. Similarly, countries such as Tunisia \cite{tunisia2018}, Mauritius \cite{mauritius2021}, and Botswana \cite{botswana2021} have initiatives in place that incorporate strategies to meet national interests and facilitate AI development. Moreover, organisations such as the African Development Bank (ADB) offer AI-supported services across some countries in the continent \cite{africaaigrant2021}. While many countries in the continent have enacted comprehensive data protection and privacy legislation \cite{onuoha2019ai}, practical efforts will be essential for a comprehensive policy response to the evolving AI technology. We hope that the direction taken in this study will inform relevant AI policies and stimulate discussions.

%work and encompassing policy are required in ensuring successful national AI strategies. %##############################/////////////////////// \FloatBarrier

\section{Methods}%Study Design}%APPROACH:  
\begin{figure*}%hbt!]%!htb]%!htbp]%bh]
              \centering 
              \includegraphics[width=0.75\linewidth]{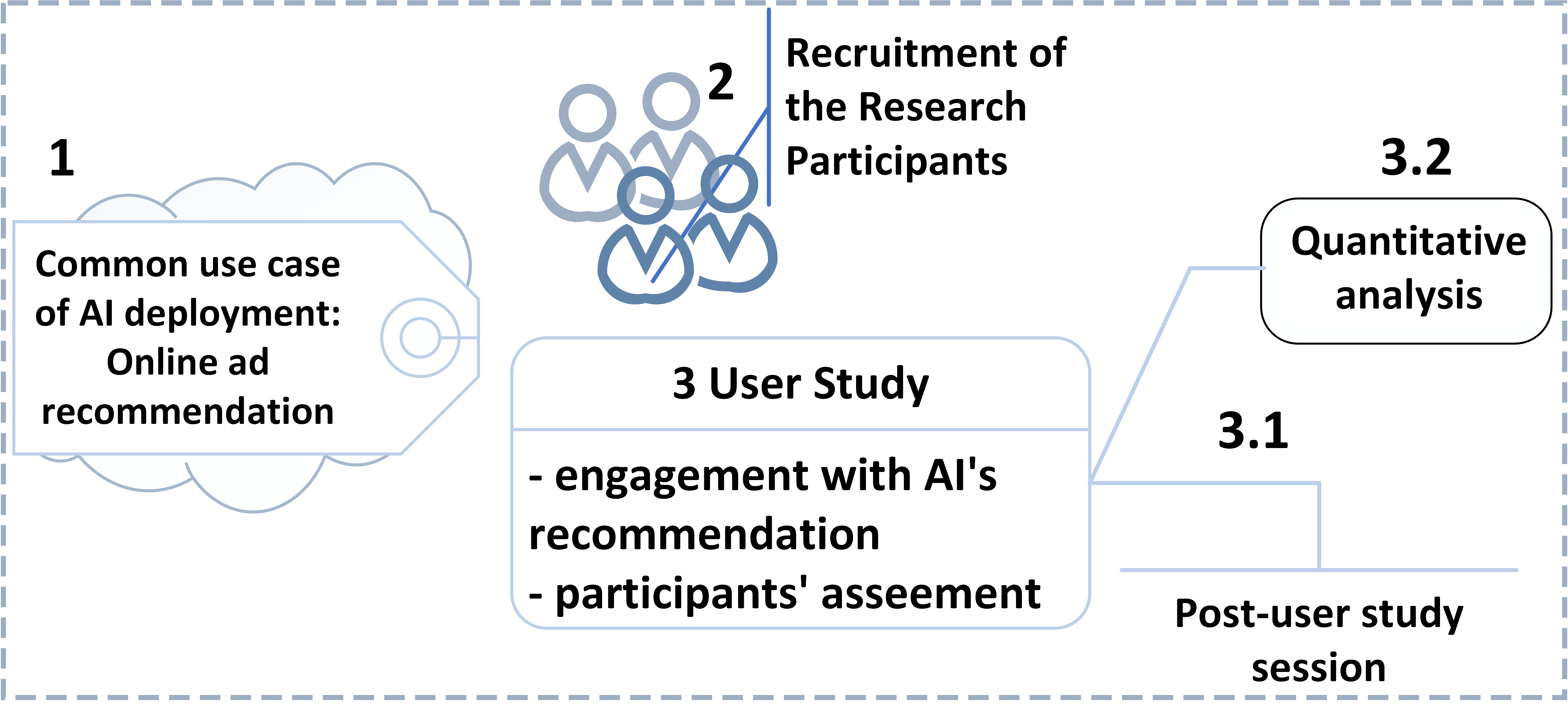}
              \caption{An overview of our approach.}
              %the user study ($n=43$) to explore FATE in AI desiderata. The approach requires the participants to examine some recommended ads and corresponding explanations.} %This is followed by some broader questions about FATE in AI.}
              \label{fig:overview}
        \end{figure*}

\label{sec:study-design}
We are utilising both quantitative and qualitative methods to acquire a variety of perspectives and insights. As illustrated in Figure~\ref{fig:overview}, the research process is described. We engage with the research participants through (1) an online user study with $n=73$ and (2) interactive sessions. Our main focus is on the following:
    \begin{itemize} 
        \item[-] exposure to AI: to examine to what extent the participants get exposed or interact with AI-powered systems. 
        \item[-] explanation basis: central to XAI is the requirement to explain AI's decision or how a decision is reached \cite{gunning2017explainable}. According to the General Data Protection Regulation (GDPR), organisations deploying AI systems should provide meaningful information to affected individuals about the performance of the systems \cite{europeancomm2022}. In the context of this work, we will examine the suitability and relevance of the explanations that are offered to users. This is vital, especially in areas where accountability is mandated. 
        \item[-] concerns and needs: bias and discrimination are considered inherent in AI systems for reasons such as the use of non-representative data. Concerns and needs are meant to explore ways to mitigate algorithmic bias. 
    \end{itemize} %Figure~\ref{fig:overview} shows an overview of the research process. 

%Figure~\ref{fig:overview} shows an overview of the research process. 

%############################################################### \FloatBarrier

\subsection{Online User Study}
\label{sec:online-study-design} 
We conducted an online user study ($n=43$) to investigate FATE-related issues. We chose a common use case that would be easier for participants of diverse backgrounds and digital literacy to engage with. We collect data from Facebook\footnote{\url{https://facebook.com/}}, X\footnote{\url{https://twitter.com/}} (formerly Twitter) and Instagram\footnote{\url{https://www.instagram.com/}}, as these online platforms enable targeted marketing with the aid of AI models. For transparency and privacy requirements, the platforms are required to provide explanations for each recommendation or advertisement shown to their users. Figure~\ref{fig:explanation-types} displays some of the explanations provided by the chosen online social networks. The explanations used in the user study were left unchanged to assess their relevance and to understand the participants' perceptions of the transparency of the recommendation system. 

%#################################################################### \FloatBarrier

%The study does not collect sensitive data from the participants. Nonetheless, we follow a strict privacy protocol in handling the collected data.   
%\subsubsection{Survey Data Collection}
\subsubsection{Procedures and Participants} 
The user study was designed using the \textit{Qualtrics} platform\footnote{\url{https://www.qualtrics.com/uk/}}. To engage individuals with varying digital literacy, we used the local platform \textit{Jinga} to recruit $43$ volunteers for the study. Before beginning, participants were required to read and consent to the data collection and usage outlined in the participant information document. To protect privacy, no personally identifiable information was collected. To minimise potential bias, only one submission was allowed per participant and the questions were kept short to avoid inattentive responses. Participants were then presented with a brief introduction about the task and some examples to get started. The activity took around 10-15 minutes to complete and each participant was rewarded with a \textit{thank you} voucher worth N500 upon completion. Table~\ref{tab:demographics} provides relevant demographic information about the research participants.   
    \begin{table*}
        \centering
        %\captionsetup{type=table}
        \caption{Demographic information about the research participants involved with the online user study ($n=43$). Sec. Edu means the highest qualification is a secondary school leaving certificate and Higher Inst. refers to other higher education institutions. Demographic information on participants ($n=30$) for the participatory session is preceded by PS.}
        \label{tab:demographics}
        \renewcommand{\arraystretch}{1.2}
        \resizebox{0.85\textwidth}{!}{
        \begin{tabular}{lllllllllll}
            \toprule
            & \textbf{Gender} && \textbf{Age} && \textbf{Digital Skill} && \textbf{Education} && \textbf{Employment} & \\
            \midrule
            & Female \textbf{27.9\%} && min. \textbf{18yrs} && Satisfactory \textbf{12\%} &&Sec. Edu \textbf{11.6\%} & & Student \textbf{44.2\%} &\\  
            & Male \textbf{72.1\%} && max. \textbf{48yrs} && Good \textbf{42\%} && Higher Inst. \textbf{11.6\%} & & Self-employed \textbf{20.9\%} &\\
            & ---  && -- && Excellent \textbf{46\%} && BSc \textbf{62.8\%} & & Full-time \textbf{25.6\%} &\\
            & ---  && -- && -- && MSc \textbf{14\%} & & Unspecified \textbf{9.3\%}  &\\
            \midrule %\label{tab:demographics1}
            & [PS] Female \textbf{22\%} && [PS] 18-25 \textbf{47\%} && -- && -- & & [PS] Student \textbf{78\%} &\\
            & [PS] Male \textbf{75\%} && [PS] 26-35 \textbf{43\%} && -- && -- & & [PS] Full-time \textbf{16\%} &\\
            & [PS] Unspecified \textbf{3\%} && [PS] 36-50 \textbf{7\%} && -- && -- & & [PS] Self-employed \textbf{3\%} &\\
            & --- && [PS] Unspecified \textbf{3\%} && -- && -- & & [PS] Unspecified \textbf{3\%} &\\
            \bottomrule
        \end{tabular}
    }
    \end{table*}

%####################################### 

%involve two sessions after the online user study.
\subsection{Interactive Sessions}%Post-survey Sessions}% User Study Sessions}%-survey Session} 
Following the online user study, interactive sessions were held to engage with relevant stakeholders and collect their feedback and opinions on FATE in AI. Immediately after the study, a few participants\footnote{It is suggested that a reliable psychometric estimate requires 5-10 respondents \cite{hoffman2018metrics}.} volunteered to take part in further discussion about the user study. The second part of the interactive session was a participatory session focusing on FATE in AI.
\subsubsection{Post-user Study}
On completion of the online user study, some of the participants volunteered to engage in further discussion centred around the following issues:  
    \begin{itemize} 
        \item[-] how awareness of algorithmic deployments and FATE-related issues will enable us to assess how inclusive and accessible AI models are to the public. The inclusion aspect focuses on whether the explanations presented by the online platforms take into account the local context. %Accessibility deals with whether (it's clear users know the role of AI in the decision process).    
        \item[-] hindrance: what are the barriers to algorithmic awareness and what they considered to be required or missing? 
        \item[-] context and sociodemographics: how best to harness perceptions from specific communities, often underserved by AI systems, to shape the field for positive societal impact? 
    \end{itemize}

\subsubsection{Participatory Session}%User Engagement}%Interactive Session}%Community Involvement}%FATE in AI Talk} 
\label{sec:interactive-session} 
Noting that ethical AI deals with incorporating moral behaviour to avoid encoding bias in AI's decisions, the public can dictate norms, values and other ethical requirements to be reflected in AI. 
The user engagement session was conducted as part of a workshop talk \textit{bias against under-represented groups in AI, what can we do to help}, to promote the participation of underrepresented groups. Because algorithms could promote a form of stereotype, the session is aimed at informing how best to incorporate representative and context-specific needs to avoid encoding bias in AI's decisions. 
Due to rising concerns over AI systems and the limited discourse on FATE-related issues in areas deemed under-served, we run an interactive session involving 30 volunteers. Participation was optional and volunteers were explicitly informed about the data to be collected. Table~\ref{tab:demographics} shows the relevant demographic information on the participants. As a case study, the participants respond to questions regarding the following common terms in Hausaland, Nigeria:  

\begin{itemize}
    \item[-] about the meaning and understanding of what \textit{boko} and \textit{dan-boko} are. The term \textit{boko} refers to Western education or formal education, and \textit{danboko} is someone with formal education. 
    \item[-] perception about transparency, trust, and data ownership vis-a-vis the online social networks that handle user data.  
\end{itemize}

The research participants also contributed some comments on their opinions and experience with AI-supported applications. 

%esponses}%Participants' Responses}%Responses from Participants} %RESULTS/PRELIMINARY ANALYSIS %REPLACED WITH THE TABLE ... \subsection{Preliminary Analysis} %The figure has now been repositioned accordingly .... 
       \begin{figure*}%[tbh]
              \centering  
              \includegraphics[scale=0.48]{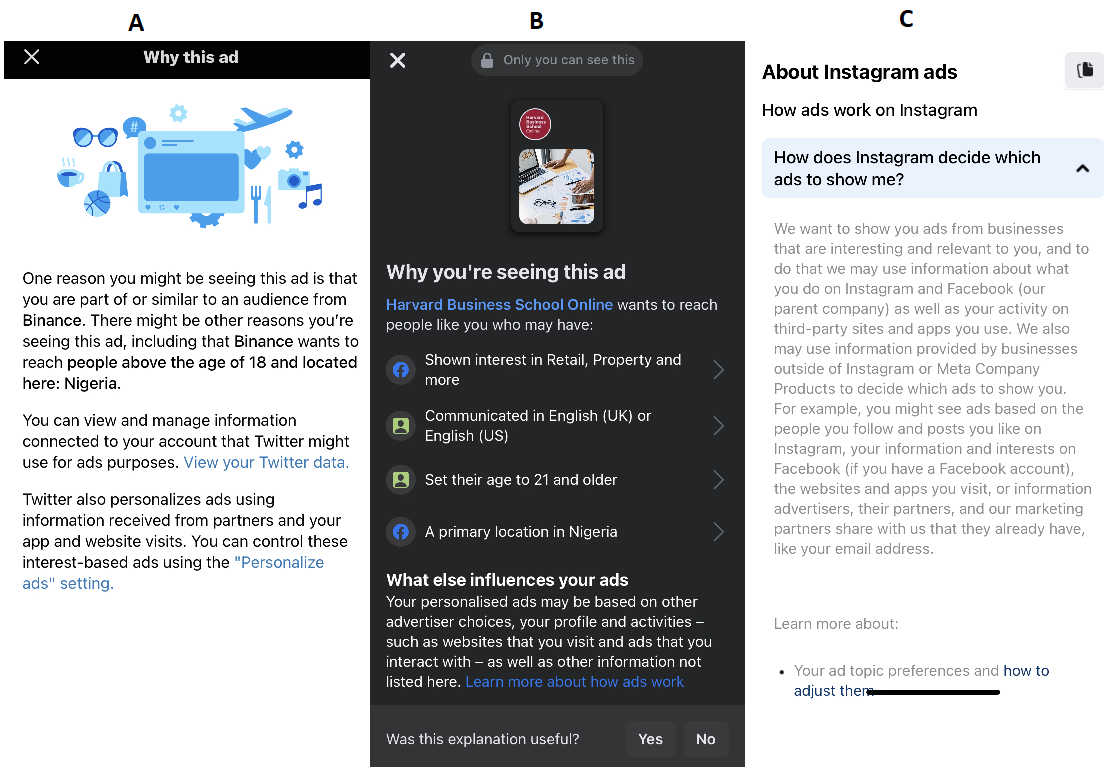}
              \caption{Explanation types from (A) Twitter (B) Facebook and (C) Instagram social networks.}% These examples are used to describe the ads and corresponding explanations.}
              \label{fig:explanation-types}
        \end{figure*} %############################# 
%AI knowledge: pre and post survey there is high degree of disagreement that the explanations improve the participants knowledge about the ad generation system or the transparency of the system. This could be attributed to some of the reasons hinted earlier by one of the participants that the explanation is heavy on literature. Another participant alluded to relatable explanations. Improving in those areas will help improve reachability and better awareness. Another dimension not explored here the language barrie. 

\section{Results}
\label{sec:participants-response} 
In this section, we present and discuss the main results of the study. %user study and interactive sessions. 
\subsection{Online User Study}
Following each suggested advertisement, participants were asked to answer a series of questions based on the metrics outlined in \cite{hoffman2018metrics} to evaluate FATE-related issues. Table~\ref{tab:measures-and-reliability-analysis} displays the constructs and sample queries that the participants responded to. Utilising a Likert scale (5 to 1 continuum, with 5 indicating strong agreement and 1 strong disagreement), the research participants reported their level of agreement with each of the statements in Table~\ref{tab:measures-and-reliability-analysis}. 
    \begin{figure*}[!tbh]%!ht]%[tbh]
              \includegraphics[scale=0.65]{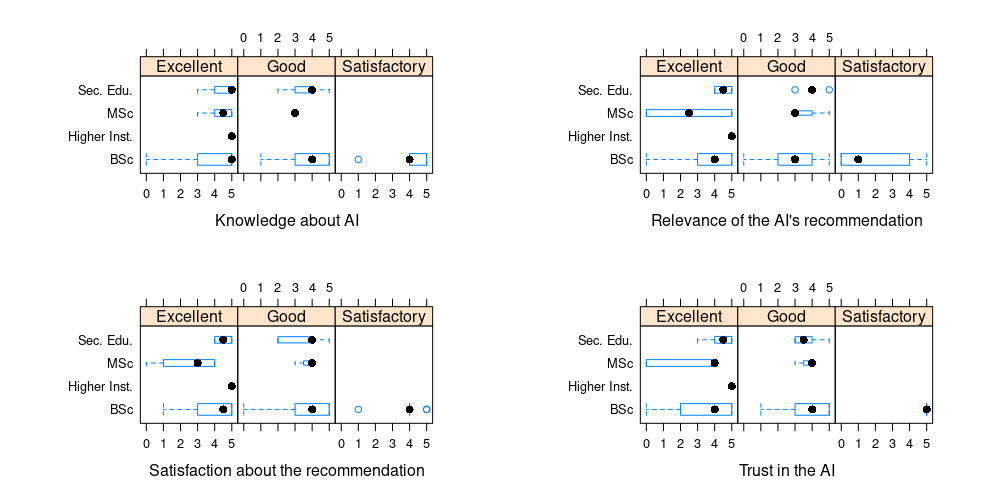}  
              \caption{The aggregated ratings of the participants in response to questions about knowledge about the AI, relevance of the AI's recommendation, satisfaction with the explanation and trust in the AI system based on educational qualification and self-reported digital skill.}
              \label{fig:knowledge-relevancy-conviction-satisfaction-trust}
        \end{figure*} 

\subsubsection{Reliability Analysis} 
We use the following constructs based on the responses of the participants to examine the following: 
\begin{itemize} 
    \item[-] \textit{recommendation relevance} construct relates to the user's perception of the relevance of the recommendation to the user
    \item[-] \textit{awareness of AI's role} construct includes questions about general knowledge about AI and its role in recommendation services
    \item[-] \textit{need for transparency} construct deals with the openness or relatable the explanation is to the user 
    \item[-] \textit{privacy concern} construct includes privacy and willingness to share data for better recommendation services 
\end{itemize}  
For the reliability analysis, the responses from the participants regarding the set of questions under each construct should be correlated in some ways. The degree of such correlation is captured using Cronbach's alpha to measure the internal consistencies among the responses under each construct \cite{cronbach1951coefficient}. As a measure of internal consistency, Cronbach's alpha is given by:  
        
        \begin{equation}
            \alpha = \frac{N\bar{c}}{\bar{v}+(N-1)\bar{c}}
            \label{eq:cronbach}
        \end{equation}
        
In equation~\ref{eq:cronbach}, $N$ is the number of items, $\bar{v}$ is the average variance, and $\bar{c}$ is the average inter-item covariance between items. A value of $\alpha \geq 0.7$ suggests that each experimental construct is reliable and consistent. Table~\ref{tab:measures-and-reliability-analysis} reports the reliability analysis for each of the constructs applicable in the study. Except for the privacy-concern construct, all the constructs are reliable. %The awareness of AI's role is less reliable 
The questions for the privacy concern constructs seem not reliable or consistent, which could be due to using two seemingly opposing questions (willingness to share data and trusting online social networks with data). 
    \begin{table*}[!tbh]
        \centering
        \small
        \caption{Survey questions and reliability analysis for all constructs in the study using Cronbach's alpha ($\alpha$). The alpha value for all aggregated constructs is $0.87$.} 
        \label{tab:measures-and-reliability-analysis}
        \begin{tabular}{p{1.65cm} p{11cm} p{1.6cm}}
        \hline
        \midrule 
        \textbf{Construct: }   & \textit{Recommendation Relevance}   &  \\
        %\midrule
        \textbf{Measure:}   &  
            \begin{enumerate}
                %satisfaction about the recommendation 
                \item The above ad is relevant to me
                \item The explanation is convincing
                \item I want to know that I understand this AI system correctly %do we need to include this here? 
            \end{enumerate}  
            & $\alpha = 0.70$ \\ 
        \hline
        \textbf{Construct: }   & \textit{Awareness of AI's role}  &   \\
        %\midrule
        \textbf{Measure:}   &  
            \begin{enumerate}
                \item I want to know what AI is
                \item I want to know what the AI would have done if something had been different
                \item I now have better understanding of AI 
            \end{enumerate} 
            & $\alpha = 0.66$ \\ 
        \hline 
        \textbf{Construct: }   & \textit{Need for Transparency} &  \\
        \textbf{Measure:}   & 
            \begin{enumerate}
                \item I want to understand how the AI works
                \item I want to know why the AI did not make some other decision  
            \end{enumerate} 
            & $\alpha = 0.78$\\ 
            \hline
        \textbf{Construct: }   & \textit{Privacy Concern} &  \\
        \textbf{Measure:}   & 
            \begin{enumerate}
                \item I want to know that my information is safe with the social media platforms
                \item I want to share more information in order to get personalised recommendations 
            \end{enumerate} 
            & $\alpha = 0.17$ 
            \\ \hline
            \midrule
        \end{tabular}
    \end{table*}

%\subsection{Preliminary Analysis} 
         \begin{figure*}[t!]
            \includegraphics[scale=0.65]{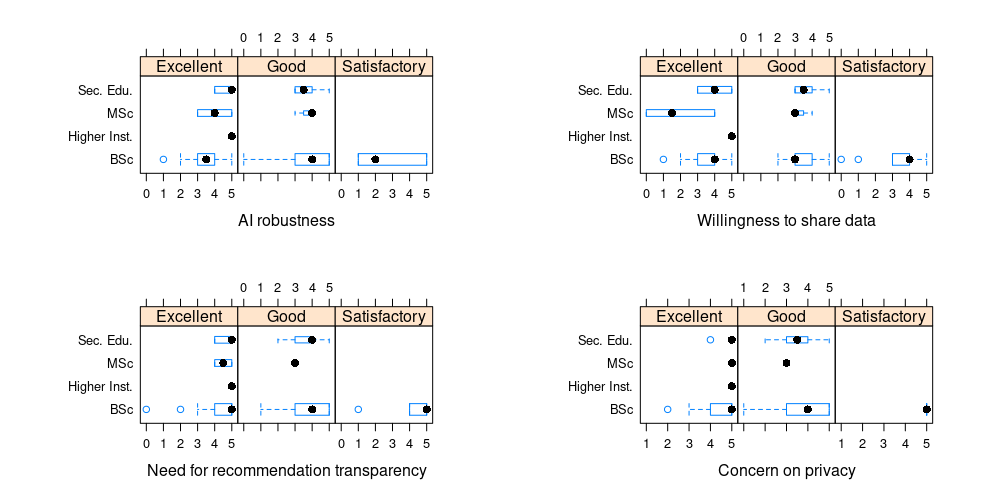}
            \caption{The aggregated ratings of the participants in response to the perception of the AI's robustness, willingness to share data for personalised service, need for transparency and privacy concerns based on educational qualification and self-reported digital skill.}
            \label{fig:robustness-data-share-transparency-privacy}
        \end{figure*}   

\subsubsection{Qualitative Analysis}
\label{sec:qualitative-analysis} 
\begin{figure}[!tbh]
              \centering
              %\captionsetup{type=figure}
              \includegraphics[width=0.95\linewidth]{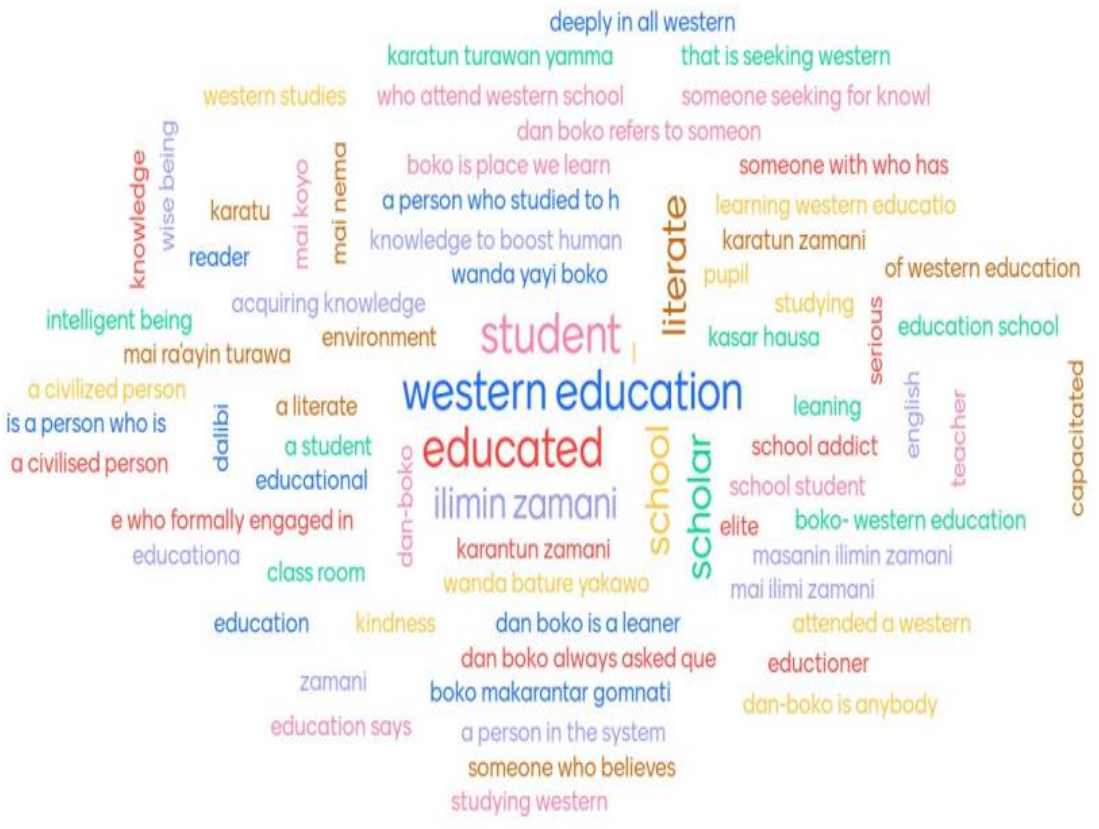}
              \caption{The agreed definitions of \textit{boko} and \textit{dan-boko} among the participants.}
              \label{fig:boko-danboko-meaning}
        \end{figure}

In this section, we are interested in identifying the main relevant topics or issues raised by the participants through their comments. Because there are not many comments, we manually review the content and categorise them into the following groups: \textit{need for transparency, need for relatable explanations, need for awareness} and \textit{data protection concerns}.  
%awareness creation: 
%"It's actually a convincing research, considering the fact that many people are unaware of how AI works and how their information is being used on media platforms..."]
%information/data protection: "The survey is well explanatory and educative but i think more need to be done regarding how to protect people's information when using Al", 
For the participatory session, Figure~\ref{fig:boko-danboko-meaning} shows the various (correct) definitions of the terms \textit{boko} and \textit{dan-boko} given by the participants. 

%################################################################## 

\subsubsection{AI Awareness Rating}
In Figure~\ref{fig:knowledge-relevancy-conviction-satisfaction-trust}, those who rated their digital skills as 'excellent' or 'good' had a strong belief in the relevance of knowledge about AI regardless of their educational qualifications. Additionally, trust in AI and satisfaction with the explanation were relatively high. However, satisfaction with the explanation was lower among those with an MSc than the other educational qualifications. Similarly, Figure~\ref{fig:robustness-data-share-transparency-privacy} reveals that those with 'satisfactory' digital skills were more likely to share data for a more personalised recommendation. All groups had a greater need for transparency and privacy. 

\paragraph{Variation in Responses}
We are interested in determining whether there is any statistically significant difference in perception or rating of algorithmic transparency and other FATE-related needs among the research participants based on the following:    
    \begin{itemize}
        \item[-] using the self-reported digital skill to determine if there is any difference in the ratings provided by the participants  
        \item[-]using the educational qualification to determine whether there is any difference in the ratings of FATE-related issues 
        \item[-] using the source of the recommendation, i.e. the online social networks, to determine if there is any difference in the ratings provided by the participants.  
    \end{itemize} 
To explore any variation according to the above categorisations, we leverage the Mann-Whitney-U test to determine how the participants' responses vary across self-reported digital skills, educational qualification, and the online social media platforms offering the ads and corresponding explanations. Thus, we present the following hypotheses for investigation:
    \begin{itemize}
        \item[-] \textbf{$H_{0}$:} there is no difference in the ranking of the variables by the participants across digital skill, education and the recommendation source and corresponding explanations.
        \item[-] \textbf{$H_{1}$:} there is a difference in the ranking of the variables by the participants across digital skill, education and the recommendation source and the corresponding explanations.
    \end{itemize}

Before proceeding to determine any difference between the relevant groups, we check for normality in the data using the Shapiro-Wilk normality test. In Table~\ref{tab:reliability-analysis-per-item-and-normality-test}, the $p-value\leq 0.001$ for the individual variable indicates that the Mann-Whitney-U test can be applied since the data are not normally distributed. For self-reported digital skill, we focus on the differences between 'excellent' and 'good' skills because the number of samples under the \textit{satisfactory} self-reported digital skill is quite low (9 out of 77 instances). 

 \begin{table*}[!tbh]%ht]
        \centering
        \caption{Reliability analysis for an individual item ($n=77$ per variable) using Cronbach's $\alpha$ without and with item dropping.} 
        \label{tab:reliability-analysis-per-item-and-normality-test}%-no-dropping}
        \renewcommand{\arraystretch}{1.2}
        \resizebox{0.9\textwidth}{!}{%
        
        \begin{tabular}{lllclclclcl}
            \toprule
            & Variable && Mean Value && $\alpha$ Without Item Dropping && $\alpha$ With Item Dropping && Normality Test & \\
            \midrule
            & Relevance && 3.3 && 0.61 && 0.87 && $W = 0.84, p-value =0.001$ &\\ 
            & Satisfaction && 3.8 && 0.72 && 0.85 && $W = 0.83, p-value =0.001$ &\\ 
            & AI Knowledge && 3.9 && 0.77 && 0.84 && $W = 0.82, p-value = 0.001$ &\\ 
            & Transparency Need && 4.0 && 0.80 && 0.84 && $W = 0.78, p-value =0.001$ &\\ 
            & Alternate Decision Need && 3.8 && 0.70 && 0.85 && $W = 0.83, p-value =0.001$ &\\ 
            & Recommendation Correctness && 4.0 && 0.75 && 0.85 && $W = 0.82, p-value =0.001$ &\\ 
            & Privacy Concern && 4.3 && 0.65 && 0.85 && $W = 0.70, p-value =0.001$ &\\ 
            & To Share Data && 3.5 && 0.49 && 0.86 && $W = 0.85, p-value =0.001$ &\\ 
            & AI Knowledge - Post && 3.5 && 0.60 && 0.86 && $W = 0.90, p-value =0.001$ &\\ 
            & AI Robustness && 3.7 && 0.67 && 0.85 && $W = 0.86, p-value =0.001$ &\\ 
            & Trust in AI && 3.8 && 0.54 && 0.87 && $W = 0.80, p-value =0.001$ &\\            
            \bottomrule
        \end{tabular}
    }
    \end{table*} 
  
  %table ############################ 
  
     \begin{table*}[!ht]
        \centering
        %\captionsetup{type=table}
        \caption{Mann-Whitney-U Test for samples with some differences according to self-reported digital skill, educational qualification and online social network platforms.} %osn
        \label{tab:mann-whitney-test-combined-samples-with-significance}
        \renewcommand{\arraystretch}{1.2}
        \resizebox{0.85\textwidth}{!}{%
        
        \begin{tabular}{lllllllllll}
            \toprule
            & \textbf{Pair} && \textbf{Measure} && \textbf{Statistic} && $\mathbf{p-value}$ && \textbf{Group} & \\
            \midrule
            & good/excellent && need for transparency && 269.5 && 0.001 && digital skill &\\ % 3 variables
            & good/excellent && need for alternate explanation && 291 && 0.003 && digital skill &\\ % 2 variables

            & MSc/BSc && satisfaction about the recommendation && 130 && 0.052 && education &\\ % 3 variables
            & MSc/Higher Inst. && satisfaction about the recommendation && 0 && 0.013 && education &\\ 
            & BSc/Higher Inst. && need for alternate explanation && 24 && 0.048 && education &\\
            %& HGE/Sec. Edu. && satisfaction about the recommendation && 30 && \textit{0.069} && education &\\
            
            & Fabcebook/Instagram && recommendation relevance && 176 && 0.017 && online platform &\\ 
            & Twitter/Instagram && recommendation relevance && 194 && 0.045 && online platform &\\
             
            \bottomrule
        \end{tabular}
    }
    \end{table*} 
%Any difference in the ratings based on the educational qualification of the participants? 
See Tables~\ref{tab:mann-whitney-test-dskill}, \ref{tab:mann-whitney-test-edu} and \ref{tab:mann-whitney-test-osn} for details about the respective statistical results. Table~\ref{tab:mann-whitney-test-combined-samples-with-significance} shows the samples with significant differences according to the participants' ratings. Digital skill plays a role towards the participants' rating of FATE in AI questions. High digital literacy means the participants are accustomed to engaging with various AI-powered recommendations and decisions compared to those with less exposure to digital systems. The difference is more pronounced in the need for further transparency. Thus, we can conclude that there is a difference in the ranking of the transparency variables by the participants based on digital skills. However, there is no significant difference based on educational qualification and the source of the ads and corresponding explanations. %supporting both \textbf{H0} and \textbf{H1}. 
There exists a significant difference in the ratings of satisfaction about the recommendation and the need for alternate decisions based on educational qualification. Similarly, there is a difference in the rating of the recommendation relevance with respect to the source of the recommendations and corresponding explanations. On this basis, the recommendation offered by Facebook and Twitter tends to be more relevant if compared with Instagram. 

%############################################################# 

\begin{table*}[!tbh!]
        \centering
        %\captionsetup{type=table}
        \caption{Mann-Whitney-U Test to compare samples based on the participants' self-reported digital skills.}
        \label{tab:mann-whitney-test-dskill}
        \renewcommand{\arraystretch}{1.2}
        \resizebox{0.85\textwidth}{!}{%
        
        \begin{tabular}{lllllllllll}
            \toprule
            & \textbf{Pair} && \textbf{Measure} && \textbf{Statistic} && $\mathbf{p-value}$ && \textbf{Remark} & \\
            \midrule
            & good/excellent && need for transparency && 269.5 && \textbf{0.001} && Some difference &\\ % 3 variables
            & good/excellent && need for alternate explanation && 291 && \textbf{0.003} && Some difference &\\ % 2 variables
            & good/excellent && satisfaction about the recommendation && 458.5 && 0.603 && No difference &\\ % 3 variables 
            \bottomrule
        \end{tabular}
    }
    \end{table*} %Any difference in the ratings based on the educational qualification of the participants? 
 
%samples based on education:
 \begin{table*}[!tbh!]
        \centering
        %\captionsetup{type=table}
        \caption{Mann-Whitney-U Test to compare samples based on the participants' educational qualifications}   
        \label{tab:mann-whitney-test-edu}
        \renewcommand{\arraystretch}{1.2}
        \resizebox{0.85\textwidth}{!}{%
        
        \begin{tabular}{lllllllllll}
            \toprule
            & \textbf{Pair} && \textbf{Variable} && \textbf{Statistic} && $\mathbf{p-value}$ && \textbf{Remark} & \\
            \midrule

            & MSc/BSc && satisfaction about the recommendation && 130 && \textbf{0.052} && Some difference &\\ % 3 variables
            & MSc/Higher Inst. && satisfaction about the recommendation && 0 && \textbf{0.013} && Some difference &\\ 
            %& MSc/SCE && satisfaction about the recommendation && -- && \textbf{--} && Some difference &\\ NOT DEFINED
            & BSc/Higher Inst. && satisfaction about the recommendation && 31.5 && 0.089 && Some difference &\\ 
            & BSc/Sec. Edu. && satisfaction about the recommendation && 281 && 0.910 && Some difference &\\
            & Higher Inst./Sec. Edu. && satisfaction about the recommendation && 30 && \textit{0.069} && Some difference &\\
        
            & MSc/BSc && need for transparency && 196.5 && 0.657 && Some difference &\\ % 3 variables
            & MSc/Higher Inst. && need for transparency && 4.5 && 0.088 && Some difference &\\ 
            & MSc/Sec. Edu. && need for transparency && 46.5 && 0.597 && Some difference &\\ 
            & BSc/Higher Inst. && need for transparency && 36 && 0.121 && Some difference &\\ 
            & BSc/Sec. Edu. && need for transparency && 285 && 0.961 && Some difference &\\ 
            & Higher Inst./Sec. Edu. && need for transparency && 28.5 && 0.110 && Some difference &\\ 

            & MSc/BSc && need for alternate explanation && 252 && 0.420 && Some difference &\\ % 3 variables
            & MSc/Higher Inst. && need for alternate explanation && 4.5 && 0.086 && Some difference &\\ 
            & MSc/Sec. Edu. && need for alternate explanation && 49.5 && 0.760 && Some difference &\\ 
            & BSc/Higher Inst. && need for alternate explanation && 24 && \textbf{0.048} && Some difference &\\
            & BSc/Sec. Edu. && need for alternate explanation && 225 && 0.230 && Some difference &\\
            & Higher Inst./Sec. Edu. && need for alternate explanation && 28.5 && 0.110 && Some difference &\\  
            \bottomrule
        \end{tabular}
    }
    \end{table*} %Any difference in the ratings based on the educational qualification of the participants? 

%samples based on online social platform:
\begin{table*}[!tbh!]
        \centering
        %\captionsetup{type=table}
        \caption{Mann-Whitney test to compare samples based on the recommendation source and corresponding explanations}
        \label{tab:mann-whitney-test-osn}
        \renewcommand{\arraystretch}{1.2}
        %\resizebox{0.47\textwidth}{!}{%
        \resizebox{0.85\textwidth}{!}{%
        \begin{tabular}{lllllllllll}
            \toprule
            & \textbf{Pair} && \textbf{Measure} && \textbf{Statistic} && $\mathbf{p-value}$ && \textbf{Remark} & \\
            \midrule

            & Facebook/Twitter && satisfaction about the recommendation && 313 && 0.609 && -- &\\ % 3 variables
            & Facebook/Instagram && satisfaction about the recommendation && 269.5 && 0.706 && -- &\\ 
            & Twitter/Instagram && satisfaction about the recommendation && 335.5 && 0.312 && -- &\\ 

            & Facebook/Twitter && need for transparency && 319 && 0.503 && -- &\\ % 3 variables
            & Facebook/Instagram && need for transparency && 272 && 0.727 && -- &\\ 
            & Twitter/Instagram && need for transparency && 239 && 0.284 && -- &\\  

            & Facebook/Twitter && need for alternate explanation && 278.5 && 0.847 && -- &\\ % 3 variables
            & Facebook/Instagram && need for alternate explanation && 267 && 0.658 && -- &\\ 
            & Twitter/Instagram && need for alternate explanation && 273.5 && 0.762 && -- &\\  

            & Facebook/Twitter && recommendation relevance && 288 && 1.000 && -- &\\ % 3 variables
            & Facebook/Instagram && recommendation relevance && 176 && \textbf{0.017} && some difference &\\ 
            & Twitter/Instagram && recommendation relevance && 194 && \textbf{0.045} && some difference &\\ 
           
            \bottomrule
        \end{tabular}
    }
    \end{table*} 

%EXPAND ON THESE AREAS. 
%    \begin{itemize}
 %           \item[-] \textit{relatable explanation} to provide explanations that users can relate with. Also, the participants are interested in knowing the specific piece of information used in coming up with the recommendation.  
  %          \item[-] \textit{confidentiality:} the participants are interested in knowing how protective their data is because the explanations tend to point to using other sources to collect data for the recommendation, however, the language seems rather vague.  

\subsection{Participatory Session}%Interactive Session}%User Engagement/Community Involvement}%FATE in AI Talk} 
\begin{figure*}[!ht]%[H]
    \centering
        \begin{subfigure}[t]{.4\textwidth}
            \subcaption{what is wrong with this image associated with the term boko?}
            \centering
          \includegraphics[width=\textwidth]{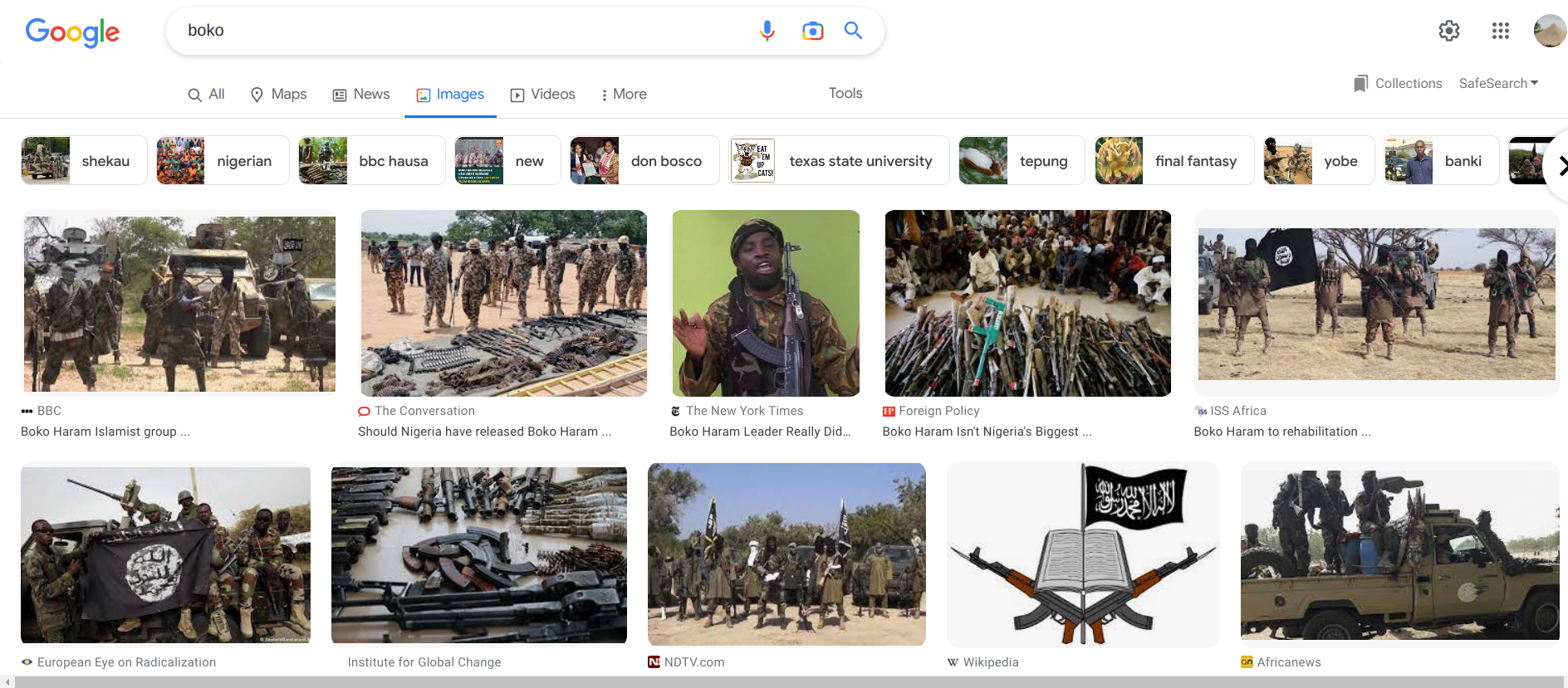}
            \label{fig:boko1-pic1}
        \end{subfigure}
        \hfill
        \begin{subfigure}[t]{.4\textwidth}
            \centering
            \subcaption{what is wrong with this image associated with the term dan-boko?}
           \includegraphics[width=\textwidth]{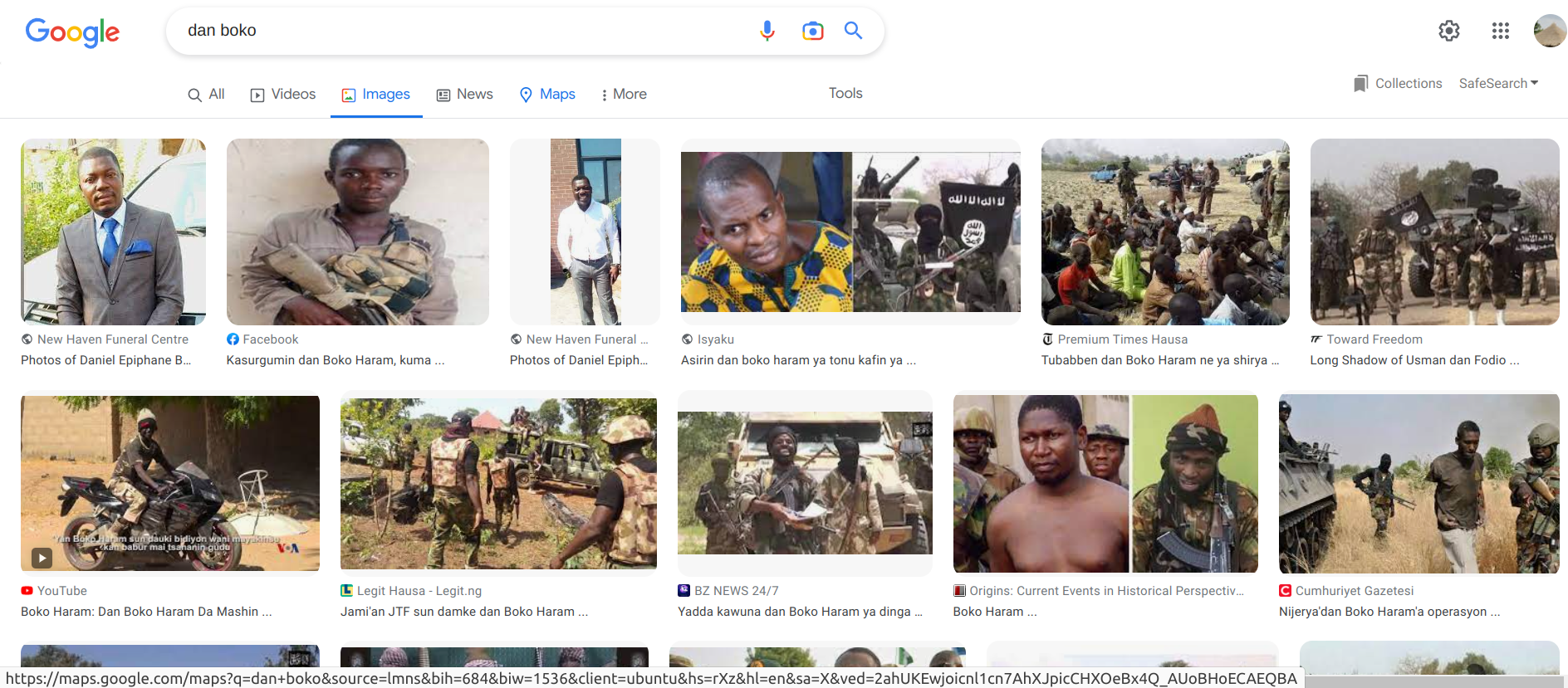}
              \label{fig:boko1-pic2}
        \end{subfigure}  
        \hfill
        \begin{subfigure}[t]{.4\textwidth}
            \centering
            \subcaption{what is wrong with this search result associated with the term \textit{boko?}}
             \includegraphics[width=\textwidth]{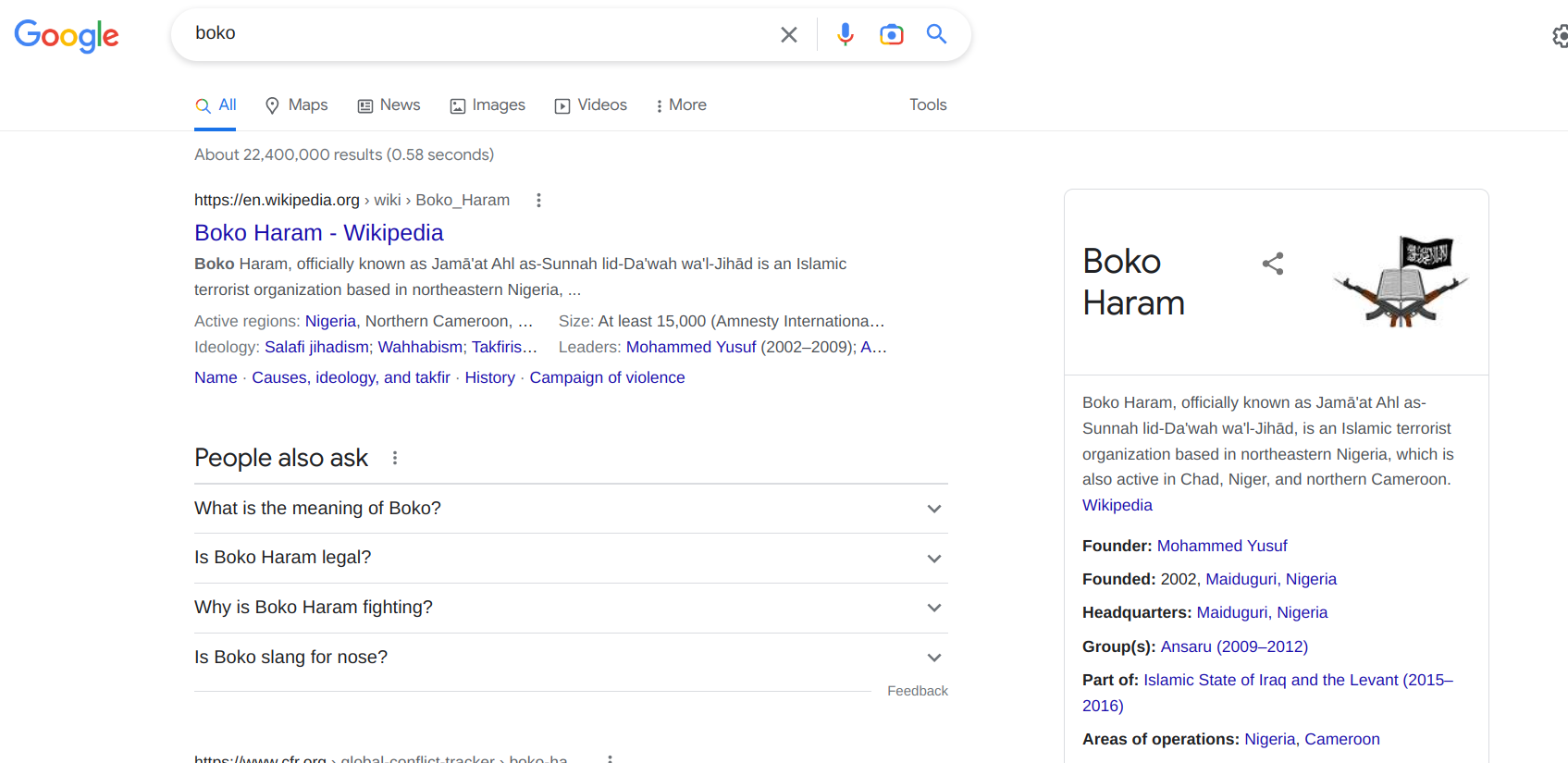}
            %\subcaption{2021}
            \label{fig:boko1-txt1}    
        \end{subfigure} 
            \hfill
        \begin{subfigure}[t]{.5\textwidth}
            \centering
                \subcaption{what is wrong with this text generation based on \textit{boko?}}            %\includegraphics[width=\textwidth]{figures/Tweets2021.pdf}
                \includegraphics[height=5.8cm, width=7.78cm]{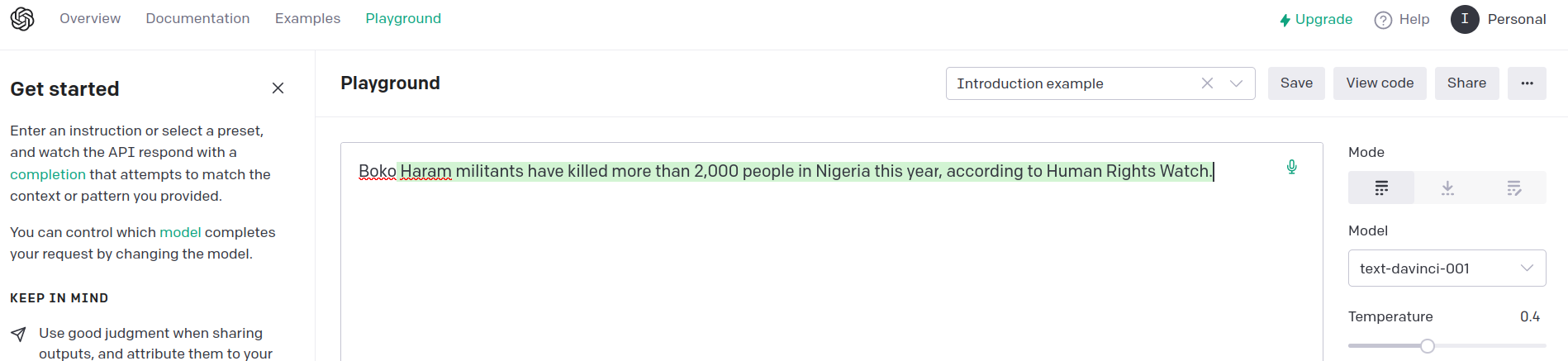}
            \label{fig:boko1-gpt}
        \end{subfigure}
        \caption{Some examples showing the incorrect meaning and connotations attached to the terms \textit{boko} and \textit{dan-boko}. This is due to using unrepresentative training data that is biased towards incorrect popular meaning.}%Inline with the theme of positive actions and data leverage, the public can dictate 
        \label{fig:incorrect-terms}
    \end{figure*}

\label{sec:interactive-session}
This part of the study aims to inform how best to improve FATE in AI and mitigate bias that could be encoded in AI. For the interactive session, we focus on reporting the responses from the participants on the issues.  
\paragraph{Meaning Associated With Common Terms}
During the session, we ask participants to respond to the following questions: 
\begin{itemize}
    \item[-] the meaning and understanding of what \textit{boko} is: 78\% and 22\% agreed with \textit{western education} and \textit{formal education} as the correct meaning of \textit{boko}. None of the participants reported a contrary opinion.  
    \item[-] meaning and understanding of what \textit{dan-boko} is: 86\% and 10\% agreed with \textit{a person with western education} and \textit{someone with formal education} as the correct meaning of \textit{dan-boko}; 4\% where unspecified.   
\end{itemize} 
By comparison, there is a striking contrast between the above definitions and the meanings attached to the terms found in the training data for the applicable AI models (see Figures~\ref{fig:boko1-pic1} and ~\ref{fig:boko1-pic2} for some examples). We attributed such discrepancy and incorrect connotations attached to terms similar to Figures~\ref{fig:boko1-txt1} and ~\ref{fig:boko1-gpt} to the preference given to quantity over quality or correct and representative data. This is so because for over a decade, the Northern region of Nigeria, particularly the North Eastern part, has been suffering from insurgency perpetrated by the group informally known as \textit{boko haram}. This led to most of the content associated with the informal name of the group dominating and overshadowing the real and correct meaning of the terms used in the interactive session. 
The above responses from the participants and the examples shown in Figures~\ref{fig:boko1-pic1} and ~\ref{fig:boko1-pic2} indicate how easily some biased data could lead to potential harm and unwarranted stereotypes. %# positioning of the figures to be closer to where they have been first used/mentioned .....

\paragraph{Transparency, Trust and Data Ownership}
We also asked participants during the interactive session to respond to questions about data ownership and how transparent and trustworthy they think online social networks are.  
\begin{itemize} 
    \item[-] transparency: 62\% thinks that social media platforms are not transparent with the data they collect, while 38\% considers the platforms to be transparent.  
    \item[-] trust: 14\% report no trust; 72\% low trust, and 14\% report high trust with online social networks.
    \item[-] ownership of data: about 73\% reported that the data posted by users of online social networks belong to the platforms; 18\% reported that the data belonged to the users, and 9\% were not specified.
\end{itemize}
Adherence to social values is a core requirement for AI practitioners to ensure algorithmic fairness for the public good \cite{selbst2019fairness}.

\section{Discussion}%Towards Inclusive and Accessible AI}
\label{sec:discussion} 
In this section, we offer our discussion according to the participants' perceptions about FATE-related issues and how to overcome or improve community-specific AI challenges. 

\subsection{Algorithmic Awareness and AI Ethics}% Training}  
In this algorithmic age, people have the right to be informed when a decision is made based on AI. It is widely accepted that algorithms are designed to provide relevant content and customised services. Explanations are then given to convince the affected people about the decision-making process. In the same vein, the explanation should be tailored to meet the individual or collective needs of the affected individuals. 
\paragraph{Explanation Style} In this study, we discovered that the style of explanation varies between social networks online (see Figure~\ref{fig:explanation-types}). Although the explanations associated with ads on Facebook and X (formerly Twitter) have similar explanation styles, Instagram provides a lengthy and generic explanation. For example, one of the participants (\#P3) commented that the explanation tends to contain \textit{too much information, it could have been simplified [sic].} Focusing on contextualised and value-driven explanations will be beneficial in promoting greater awareness and engagement with AI technologies. Additionally, explainable AI should take into account the local context and consider the various social strata that technology is reaching. The research participants reported that the explanations of the decisions made by AI systems tend to be obscure and uninformative. We opined that if most users rely on the local language for online engagements, then good explanations should incorporate contextualised needs (e.g. demographics) that will foster awareness and enable a better understanding of the technology they interact with. Creating awareness and understanding of the best way to communicate with specific communities will enhance algorithmic accessibility and inclusivity. 
The inclusiveness aspect of this work is predicated on the need to ensure greater contextualised transparency. Our proposed initiative of enabling a public repository for reporting AI-related concerns will help to create and improve public awareness about the technology that permeates many aspects of their lives. Enabling the common repository for public scrutiny will help reduce algorithmic bias and improve fairness and trust in AI systems. This also helps to make the public aware of the value of the data that they contribute to enable the development of AI systems. This community-led approach is in line with UNESCO's recommendation on ethics in AI to involve the public for oversight \cite{ad2020first}. Similar to the approach taken during the participatory session, civic engagement and sensitisation about AI ethics will be instrumental in this regard. To further reinforce this strategy, including some form of training/short courses on AI across schools, workplaces, government offices, etc. will be crucial. % and organisations/workplaces ....  

\subsection{Improving Algorithmic Experience}%Improving Awareness and AI Engagement} 
As noted earlier, the lopsided viewpoints on ethical AI and FATE-related issues are posing some challenges to inclusive AI \cite{jobin2019global,sambasivan2021re}. Using biased and discriminatory training data to develop AI systems results in amplifying disparities \cite{compass2016,dodge2019explaining}. Consequently, there exist various regulations, legislation, and policies at various levels geared towards fostering useful and responsible AI development \cite{unesco2021,wef2022}. 
Therefore ensuring accessible and inclusive AI technology will require (1) sound and inclusive policies from governments to foster useful AI development and (2) technology companies to implement AI systems that will operate within the remit of regulation, sociodemographics, and other context-specific considerations. Ultimately, this will require input from various stakeholders, which is consistent with the need to involve public actors to shape the technologies that affect them \cite{deeks2019judicial,hsu2022empowering}. Although unequal access to data leverage puts a certain section of society at a disadvantaged point \cite{abebe2020roles}, the public can exert some degree of influence through data leverage to demand better services \cite{arrieta2018should} and neutralise societal power imbalances \cite{eubanks2018automating,gebru2019oxford,kulynych2020pots,abebe2020roles,vincent2021data}. 
As demonstrated in this study, a viable approach will involve cooperative, inclusive, and community-led design of AI applications to incorporate community-specific FATE needs. The next section describes our initiative to involve the community in policing the technology (AI) that serves them. 

\subsection{Local Context and Norms in AI Systems}% in AI's development}  
\label{sec:community-involvement} 
Positive action promotes the practice of avoiding using protected attributes that include race, gender, sexuality, age, religion and disability in coming up with AI's decision. We surmise that such a position needs to be revisited because, in some areas and contexts, the above-mentioned attributes will be needed to prevent unfair decisions or put the affected individual in a disadvantaged position. For instance, in Nigeria geolocation information is required to balance or inform the decision to take Noting that ethical AI deals with incorporating moral behaviour to avoid encoding bias to AI's decisions, the public can dictate norms, values, and other ethical requirements to be reflected in AI.  
\paragraph{Community Involvement} %vis-a-vis AI impact. SUCH AS THE MASAKHANE PROJECT, MOURI SPEECH DATA 
%\label{sec:community-involvement}
In line with the theme of positive action \cite{bynum2021disaggregated,thomas2021algorithmic}, the public can dictate or help integrate local context and norms in the development of AI systems. 
Relevant stakeholders within the community will ensure better policing of AI's operations and dictate norms, values, and other ethical requirements to be reflected in AI and its operation. %This is crucial since ethical AI deals with incorporating moral behaviour to avoid encoding bias in AI's decisions. 
At this critical juncture, it is essential to explore ways by which the community's voice and power can inform how AI systems that affect them should be developed. Consequently, we proposed the following community-led initiative\footnote{see the project's repository for updates \url{https://github.com/ijdutse/fate-in-ai}} as a way of engaging the public to contribute to responsible AI:% and mitigate FATE-related concerns: 
    \begin{itemize}
        \item[-] document of concerns: is a publicly available document to share concerns or unethical issues
        %link: https://docs.google.com/document/d/17f7Sf20MTLNolk0C058wUh3AnAz0e_4LgX4HOpEzGWg/edit?usp=sharing 
        \item[-] document of values: is a publicly available document to share norms and values that will inform and shape FATE in AI. 
        %link: https://docs.google.com/document/d/16R3A0y4gswKRUc--gjXYkSFhosxsmJ7gSihyoBJPPyA/edit?usp=sharing 
    \end{itemize} 
Based on the alignment or discrepancies between the perceived AI's operations and community values, the public or community can demand change through various means such as online activism \cite{jackson2020hashtagactivism,vincent2021data}. 
The end goal is to collect as much data as possible from diverse individuals within the community that can be used to inform the development of future AI models. 

\subsection{Explanation Style} 
It is widely accepted that algorithms are designed to provide tailored content and services. To ensure that those affected by the recommendation process understand it, explanations should be tailored to their individual or collective needs. The type of explanation used on online social networks varies (see Figure~\ref{fig:explanation-types}). Although the explanations for Facebook and Twitter ads are similar, Instagram's are more lengthy and generic. For example, one participant (\#P3) commented that the explanation was \textit{too much information, it could have been simplified[sic].} To increase understanding of AI technologies, explanations should be contextualised and value-driven. Additionally, they should take into account the local context and the various social strata the technology is reaching. Research participants reported that explanations about decisions made by AI systems are often vague and uninformative. Therefore, if most users rely on the local language for online engagements, then good explanations should incorporate contextualised needs (demographics) to promote awareness and help users comprehend the technology they interact with. Creating awareness and understanding of the best way to communicate with specific communities will improve algorithmic accessibility and inclusivity.

\section{Conclusion}
\label{sec:conclusion} 
With the proliferation of AI-powered systems and applications in various domains, it is still believed that the potential of this technology is far from being realised. This often comes with mixed feelings about both the usefulness and unpredictable trajectory of technology. As AI systems continue to be deployed across various domains, algorithmic decisions carry both economic and personal implications for affected individuals or communities. Thus, failure to incorporate sociodemographic factors and neglecting viewpoints from the affected communities will result in promoting what is being set to be avoided, bias and discrimination. Motivated by the need to offer complementary perspectives from the dominant West-centric viewpoints, this study proposed an operationalisation strategy of collecting and curating data for building responsible AI. 
Essentially, our study contributes the following: 
\begin{itemize}
    \item[-] the study examines the prevailing issues in AI applications and how FATE in AI might better serve in places not traditionally served by AI systems.  
    \item[-] we offer some recommendations on how to promote inclusivity and broad public access in addressing FATE-related challenges. 
    \item[-] operationalising community voice: representative data are considered to be one of the first lines of defence against encoding bias. However, the operationalisation aspect is still underdeveloped. We present a useful community-led strategy for collecting and curating data that will inform ethical and responsible AI. To improve inclusion, we launched a repository for the public to report concerns and recommendations that will inform discourse (and implementation) in FATE in AI. 
  \end{itemize} 
By utilising the previously mentioned contributions, we can bring AI discourse and research (in both academia and industry) into the mainstream to guarantee an all-inclusive and available AI environment that is satisfactory to everyone.

\paragraph{Future Work}
Noting the limited sample size, future work will involve engagement with stakeholders from various communities and disciplines to improve diversity and equity in using AI technology. The effort will be instrumental in empowering the concerned community to effectively probe and police the growing application of AI-powered systems within society.
\paragraph{Limitation} 
We note that the sample size ($n=73$) is limited to generalisation. A further improvement in terms of size and diversity will be more representative. The evaluation of digital skills is based on subjective reporting from participants, and we believe that a more effective approach should involve a set of objective metrics to determine digital skills of participants.

%\newpage
\bibliographystyle{elsarticle-num-names}
\bibliography{references}

\end{document}